\documentclass[aps,pra,twocolumn,showpacs,amsmath,amssymb,superscriptaddress]{revtex4}
\usepackage{graphicx}
\usepackage{bbm}
\usepackage{hyperref}
\usepackage{times}
\usepackage{amsmath}
\usepackage{amssymb}
\usepackage{amsfonts}
\usepackage{epsfig}
\usepackage{color}
\usepackage[normalem]{ulem} %% \sout{text}
\usepackage{cancel}

\newcommand{\bi}{\bibitem}
\newcommand{\be}{\begin{equation}}
\newcommand{\ee}{\end{equation}}
\newcommand{\ba}{\begin{eqnarray}}
\newcommand{\ea}{\end{eqnarray}}

\newcommand{\rf}[1]{(\ref{#1})}

\begin{document}
\title{On the time optimal thermalization of single mode Gaussian states}

 \author{Alberto Carlini}
 %\email{}
  \affiliation{Universita' del Piemonte Orientale, DISIT, Alessandria, Italy}
 \affiliation{NEST, Istituto di Nanoscienze-CNR, Pisa, Italy}
\author{Andrea Mari}
 %\email{}
 \affiliation{NEST, Istituto di Nanoscienze-CNR, Pisa, Italy}
  \affiliation{Scuola Normale Superiore, Pisa, Italy}
 \author{Vittorio Giovannetti}
% \email{}
 \affiliation{NEST, Istituto di Nanoscienze-CNR, Pisa, Italy}
  \affiliation{Scuola Normale Superiore, Pisa, Italy}
 %\date{July 2, 2014}
\begin{abstract}

\centerline{ABSTRACT}
We consider the problem of time optimal control of a continuous bosonic quantum system subject to the action of a Markovian dissipation.
In particular, we consider the case of a one mode Gaussian quantum system prepared in an arbitrary initial state and which relaxes to the steady state due to the 
action of the dissipative channel.  We assume that the unitary part of the dynamics is represented by Gaussian operations which preserve the Gaussian nature of the quantum state, i.e. arbitrary phase rotations, bounded squeezing and unlimited displacements. In the ideal ansatz of unconstrained quantum control (i.e.\ when the unitary phase rotations, squeezing and displacement of the mode can be performed instantaneously), we study how control can be optimized for  speeding up the relaxation towards the fixed point of the dynamics and we analytically derive the optimal relaxation time. 
Our model has potential and interesting applications to the control of modes of electromagnetic radiation and of trapped levitated nanospheres.
 \end{abstract}  


\maketitle

\section{Introduction}

Quantum optimal control theory is by now a well established area of research with several applications in quantum information (for reviews see, e.g., Refs. [1-6].
%\cite{brif, bonnardchyba}.
A particular example of quantum control is time optimal control.
Here the aim is to determine  the optimal control strategy for a quantum system such that a given task is obtained in the minimum amount of time.
Time optimal control has been studied in a large variety of settings and perspectives, e.g. to give a more physical meaning to the complexity of quantum algorithms \cite{schulte} and to solve complex problems as geodesic evolutions in a given geometry \cite{nielsen}.
Geometrical approaches \cite{khaneja} and variational principles for constrained Hamiltonians \cite{carlini06} have been introduced, upper bounds for the speed of evolution of quantum systems in the Hilbert space (the `quantum speed limit', or QSL) have been discussed \cite{margolus98,caneva09}.
The case when the quantum system is not perfectly isolated from the environment and it is subject to decoherence \cite{breuer02} has also been extensively studied (see, e.g., Ref. \cite{roloff,bonnardsugny}). The time optimal control of qubits in dissipative environments is discussed in [16-25],
%\cite{sugny,carlini08}, 
while results related to the QSL are given in \cite{campo13}.
Recent applications of time optimal control to quantum thermodynamics can also be found in [27-29].
%\cite{kosloff,salamon}.

In this work we focus on continuous single mode systems evolving according to a Gaussian dynamics \cite{braunstein, weedbrook, hw, ew}. This model is particularly suitable for describing modes of electromagnetic radiation \cite{Qoptics}, but also different systems such as nano-mechanical resonators \cite{OM1, OM2}, trapped dielectric particles \cite{sphere1,sphere2,sphere3}, {\it etc.}.  

If a bosonic mode is in contact with a thermal (and possibly squeezed) environment, the state will naturally tend towards a constant steady state in equilibrium with the bath. This evolution corresponds to a generalized Gaussian dissipative channel \cite{hw,ew} and the asymptotic steady state is the fixed point of the dynamical map. In general the system will converge close to the fixed point in a given amount of time which depends on the initial state and on the particular model of the channel. The goal of this work is to study how quantum control can be used in order to speed up the relaxation time of the system, with respect to its natural dissipative evolution without control. In particular we consider a control composed of a sequence of unitary Gaussian operations, i.e. arbitrary phase rotations, bounded squeezing and unlimited displacements. We also assume that such operations can be applied instantaneously, meaning that during the application of the control one can neglect the dissipation induced by the environment. Furthermore, we do not allow any feedback in our system.

Even in the limit of fast and unconstrained control, the optimal control strategies and the optimal relaxation times are nontrivial. The reason is that a unitary control cannot change the purity of the state and the only possibility to reach the fixed point of the channel is a proper combination of the roles of the external Hamiltonian control and the intrinsic dissipative action of the environment. 
A similar approach was used in the case of a discrete, finite dimensional quantum system represented by a qubit \cite{mukherjee}.  In this work the analysis is extended to the case of a continuous variable single-mode Gaussian state.

In addition to the derivation of quantitative expressions for the optimal relaxation times, we also find some general results which are analogous to those obtained for a single qubit \cite{mukherjee}. Specifically, in the case of a single mode thermal state which is placed in contact with a bath at a different temperature, we find that: quantum control cannot be used to speed up the cooling rate of the system but is instead advantageous for heating up the system. Finally we have also studied the opposite task of avoiding the spontaneous thermalization of the Gaussian state and we found, similarly to the case of a qubit, that the dynamics of some particular states can be blocked by quantum control for an indefinite amount of time. In other words  
quantum control can artificially create a new set of fixed points which are far away from the natural equilibrium state.

The paper is organized as follows: in Section II we review the main properties of continuous variable quantum systems with $n$ bosonic degrees of freedom. 
Section III is devoted to the introduction of $n$-mode Gaussian quantum states and their representation in the real symplectic phase space, which is given entirely 
in terms of the displacement vector and the 
covariance matrix (CM).
In Section IV we introduce the Markovian dissipative channels which preserve the Gaussian nature of quantum states.
In particular, Section IV.A describes the unitary part of the Gaussian channel, given by a Hamiltonian which is a quadratic polynomial in the Gaussian mode quadratures, and which thus includes phase rotations, squeezings and displacements.
On the other hand, in Section IV.B we describe the nonunitary part of the Gaussian channel, represented by amplitude damping and classical Gaussian noise, and
we derive the master equation describing the dynamics of the displacement vector and of the CM.
In Section V we restrict our attention to the case of one-mode Gaussian systems and, resorting to Williamson's theorem, we parameterize the CM in terms of the purity,
the phase and the squeezing of the mode.
In Section VI we give the fixed point of the dynamics and we review the solution of the master equation for the CM of the one mode Gaussian system in the 
dissipative Gaussian channel.
In Section VII we introduce the problem of controlled time optimal evolution up to an arbitrarily small distance from the target.
In particular, in Section VII.A we analytically study how optimal control can speed up the relaxation of the mode in the case of unconstrained coherent control,
while the situation in which the control slows down the relaxation is treated in Section VII.B.
Finally, we provide some discussion of the results in Section VIII. 
In Appendix A we explicitly derive an analytical formula for the trajectories in the absence of external control.
Finally, Appendix B  compares the free and controlled dynamics for the case in which the fixed point is a pure state.

\section{Bosonic quantum systems}

We consider a continuous variable quantum system represented by $n$ bosonic modes which may correspond, e.g., to the $n$  quantized radiation modes of the electromagnetic field \cite{Qoptics}.
In the following, we adopt the notation used in \cite{weedbrook}.
To each mode $i$ we associate a Hilbert space ${\cal H}_i$ and a pair of annihilation and creation operators ${\hat a}_i, {\hat a}_i^\dagger$, such that 
$[{\hat a}_i, {\hat a}^\dag_j ] =\delta_{ij}$, for $i, j=1...n$, and we have introduced the commutator $[\hat A, \hat B]:= \hat A \hat B - \hat B \hat A$.
The total Hilbert space is thus ${\cal H}^{\otimes n} =\otimes_{i=1}^n {\cal H}_i$.
Collecting the bosonic operators together, we can define the vector $\vec{\hat b}:= ({\hat a}_1, {\hat a}_1^\dagger, ... ,{\hat a}_n, {\hat a}_n^\dagger )^\top$ whose components satisfy the commutation rules
\ba
[{\hat b}_i, {\hat b}_j ] =\Omega_{ij} ~~;~~i, j=1...2n,
\ea
with the antisymmetric symplectic form  
\ba
\Omega := \oplus_{i=1}^n \omega = 
\left ( \begin{array}{cccc} 
\omega &&&\\
& . &&\\
&& .&\\
&&&\omega\\
\end{array}\right )~~;~~\omega := \left ( \begin{array}{cc}
0&1\\
-1&0 \end{array}\right ).
\label{simplectic}
\ea
We then introduce the quadrature field operators in their Cartesian decomposition
\ba
{\hat q}_i &:= &  {\hat a}_i + {\hat a}_i^\dagger ,\nonumber \\
{\hat p}_i &:= &  -i({\hat a}_i - {\hat a}_i^\dagger ),
\ea
and arrange these into the vector
\ba
\vec{{\hat x}} :=({\hat q}_1,  {\hat p}_1, ... , {\hat q}_n,  {\hat p}_n)^\top,
\label{quad}
\ea 
whose components satisfy the commutation relations 
\ba
[{\hat x}_i, {\hat x}_j ] =2i\Omega_{ij} ~~;~~i, j=1...2n.
\label{commutation}
\ea
Throughout the paper we work in the units where $\hbar =1$.

\section{Gaussian states}

A Gaussian state is a continuous variable quantum state with density operator $\hat \rho$  which can be characterized entirely in terms of the first and second statistical moments of $\hat\rho$, i.e. in terms of the displacement vector $\vec{d}\in {\mathbbm R}^{2n}$, with components
\ba
d_i&:=&\langle {\hat x}_i\rangle =\mathrm{Tr} [{\hat x}_i\hat\rho ]~~;~~i=1...
%{\color{blue}2}
n
\label{displacement}
\ea
and of the real and symmetric $2n \times 2n$ covariance matrix (CM) $\sigma$ with components
\ba
\sigma_{ij}&:=& \frac{1}{2}\mathrm{Tr}[\{{\hat x}_i -d_i, {\hat x}_j - d_j\}\hat\rho] ~~;~~i, j=1...
%{\color{blue}2}
n,
\label{cm}
\ea
where we have introduced the anticommutator $\{\hat A, \hat B \}:= \hat A \hat B + \hat B \hat A$.
It can be shown \cite{simon} that the Heisenberg uncertainty principle for Gaussian states expressed in terms of the CM and of the symplectic form $\Omega$ 
becomes \cite{simon}
\be
\sigma + i\Omega \geq 0,
\label{rs}
\ee
which also implies positiveness of the CM, i.e. $\sigma >0$.

Any quantum state $\hat\rho \in{\cal H}^{\otimes n}$ can also be described in phase space  (real symplectic space) in terms of a quasi-probability (normalized but in general non positive) distribution, the Wigner function
\ba
W(\vec{x}):= \int_{ {\mathbbm R}^{2n}}\frac{d^{2n}\vec{\xi}}{(2\pi)^{2n}}\exp[-i\vec{x}^\top\Omega\vec{\xi}]\chi_{\rho}(\vec{\xi}),
\label{wigner}
\ea
where $\vec{\xi}\in {\mathbbm R}^{2n}$, 
$\vec{x}\in {\mathbbm R}^{2n}$ are the eigenvalues of the quadratures \rf{quad},
\ba
\chi_\rho(\vec{\xi}):={ \mathrm{Tr}} [\hat\rho \hat{D}(\vec{\xi})]
\label{chi}
\ea
is the symmetric characteristic function and 
\ba
{\hat D}(\vec{\xi}) := \exp (i\vec{\hat x}^\top\Omega\vec{\xi})
\label{weyl}
\ea
is the Weyl (displacement) operator.
In this representation, Gaussian states are defined as those bosonic states whose Wigner function is a Gaussian, i.e. 
\ba
W_G(\vec{x}):= \frac{\exp[-(1/2)(\vec{x}-\vec{d})^\top\sigma^{-1}(\vec{x}-\vec{d})]}{(2\pi)^{2n}\sqrt{\mathrm{Det} \sigma}},
\label{wignerg}
\ea
or whose characteristic function is
\ba
\chi_G(\vec{\xi}):=\exp[-(1/2)\vec{\xi}^\top(\Omega\sigma\Omega^\top)\vec{\xi}-i(\Omega\vec{x})^\top\vec{\xi}].
\label{chig}
\ea
From the latter, one can see that the square root of the CM and the displacement play the role of, respectively, the width and the center of 
the Gaussian in phase space.

Finally the scalar product between two operators $\hat O_1, \hat O_2$ can be evaluated as a scalar product between the respective characteristic functions
\ba
{ \mathrm{Tr}}[\hat O_1^\dagger \hat O_2]= \int_{ {\mathbbm R}^{2n}}\frac{d^{2n}\vec{\xi}}{(\pi)^{2n}}\chi^\ast_{O_1}(\vec{\xi})\chi_{O_2}(\vec{\xi}).
\label{scalprod}
\ea

\section{Gaussian channels}

In this paper we will only consider quantum operations (channels) which preserve the Gaussian nature of the states
\cite{hw}.
\subsection{Unitary evolution}
It can be easily shown that the unitary operations $\hat U$ which preserve Gaussianity are generated by a Hamiltonian $\hat H$ which is a quadratic polynomial 
in the creation and annihilation operators $\vec{\hat a}:=({\hat a}_1, ... {\hat a}_n)^\top$ and $\vec{\hat a}^\dagger:=({\hat a}_1^\dagger, ... 
{\hat a}_n^\dagger)$, i.e. $\hat U=\exp(-i\hat H)$ with
\ba
\hat H= i(\vec{\hat a}^\dagger\vec{\alpha}+\vec{\hat a}^\dagger F\vec{\hat a} +\vec{\hat a}^\dagger G\vec{\hat a}^{\dagger\top})+ {\mathrm H.c.},
\label{quadham}
\ea
where $\vec{\alpha}\in {\mathbbm C}^N$ and $F, G$ are $n\times n$ complex matrices.
Alternatively, in phase space the Gaussian unitaries are equivalent to an affine map
\ba
(S, \vec{d}): \vec{\hat x}\rightarrow S\vec{\hat x} +\vec{d},
\label{affmap}
\ea
where 
%{\color{red} 
$S$
%}
%: {\mathbbm R}^{2n} \to {\mathbbm R}^{2n}$} 
is a symplectic transformation which satisfies the condition
\ba
S\Omega S^\top =\Omega.
\label{simpcond}
\ea
\subsection{Dissipative evolution}

The propagation of an $n$ mode  Gaussian state in a noisy and dissipative channel where each mode is coupled with an (a priori different and uncorrelated) Markovian environment modeled by a stationary continuum of oscillators, can be described \cite{serafini2} (in the interaction picture) by the following master equation
\ba
\dot{\hat\rho} = -i[\hat H,\hat\rho ] +\mathcal{L}(\hat\rho),
\label{master}
\ea
where $\dot{\hat\rho} :=\partial \hat\rho/\partial t$,  $\gamma_i$ are the decoherence rates and the first term on the left hand side with the Hamiltonian $\hat H$ represents  the unitary part of the dynamics, while the second term $\mathcal{L}$ gives the dissipative part of the dynamics. 
The latter is explicitly written as
\ba
{\mathcal{L}}(\hat\rho)&:=&\sum_{i=1}^n\gamma_i [ (N_i+1)\tilde{\mathcal{L}}\left({\hat a}_i\right ) +N_i\tilde{\mathcal{L}}({\hat a}_i^\dagger) \nonumber \\
&-&M_i^\star \tilde{\mathcal{D}}\left ({\hat a}_i\right ) + M_i\tilde{\mathcal{D}}({\hat a}_i^\dagger )]\hat \rho, 
\label{diss}
\ea
 and $\tilde{\mathcal{L}}(\hat O)$ and $\tilde{\mathcal{D}}(\hat O)$ are the Lindbladian superoperators 
 \ba
\tilde{\mathcal{L}}({\hat O}) \hat\rho&:=& {\hat O} \hat\rho {\hat O}^\dagger - \frac{1}{2} \{ {\hat O}^\dagger {\hat O}, \hat\rho \},
\label{dissipator}\\
\tilde{\mathcal{D}}({\hat O})\hat\rho &:=& {\hat O}\hat\rho {\hat O}  - \frac{1}{2} \{ {\hat O}^2, \hat\rho \}.
\label{classicalnoise}
\ea
The terms $\tilde{\mathcal{L}}({\hat a}_i)$ and $\tilde{\mathcal{L}}({\hat a}_i^\dagger)$ represent a generalized amplitude damping, while the terms 
$\tilde{\mathcal{D}}({\hat a}_i)$ and $\tilde{\mathcal{D}}({\hat a}_i^\dagger)$ represent phase dependent fluctuations.
The coefficients $N_i\in {\mathbbm{R}}$ are the effective bath occupation numbers (related to the correlation functions via 
$\langle {\hat B}_i^\dagger(0){\hat B}_i(\nu)\rangle:=N_i\delta(\nu)$, computed over the state of the environmental bath, with operator modes 
$\hat B_i$ 
and frequency $\nu$) and
 the coefficients $M_i\in {\mathbbm{C}}$ are the squeezing parameters of the bath (related to the environment correlations via $\langle {\hat B}_i(0){\hat B}_i(\nu)\rangle :=M_i\delta(\nu)$). When $M_i\not =0$, the bath is said to be squeezed. It can be shown that, in order to generate a completely positive dynamics, the parameters should satisfy 
  \ba
 N_i(N_i+1)\geq |M_i|^2.
 \ea
Introducing linear combinations of the canonical operators, ${\hat c}_i := C_{i j}{\hat x}_j$, where $i, j =1, ... 2n$, one can also conveniently rewrite the dissipator in Eq. \rf{master} as
\ba
\mathcal{L}(\hat\rho)= \sum_{i=1}^{2n}\left [{\hat c}_i \hat\rho {\hat c}_i^\dagger - \frac{1}{2} \{ {\hat c}_i^\dagger{\hat c}_i, \hat\rho \}\right ].
\label{dissipator1}
\ea
In this way, the master equation \rf{master} describing the quantum dynamics of an $n$ mode  Gaussian system going through a Gaussian channel can be more compactly rewritten in the symplectic space representation as
\ba
\dot {\vec{d}}&=& A\vec{d}\label{masterd} \\
\dot{\sigma}&=& A\sigma +\sigma A^\top +D,
\label{mastercm}
\ea
where $A:= \Omega [H_0+2{\mathrm{Im}}(C^\dagger C)]$ is the {\it drift} matrix, $D:=4\Omega [{\mathrm{Re}}(C^\dagger C)] \Omega^\top$ is the {\it diffusion} matrix,
and the Hamiltonian is $\hat H=({\vec{\hat x}~}^\top H_0\vec{\hat x}~)/4$. 
Thus, the time-independent matrices $A$ and $D$ completely determine the dynamics of the Gaussian state.
Furthermore, from Eq. \rf{masterd} it is easily seen that the dynamics of the displacement vector decouples and given by 
\ba
\vec{d}(t)=e^{At}\vec{d}(0).
\label{dsol}
\ea

\section{One mode Gaussian states}
 
We now focus our analysis to the case of a single mode Gaussian quantum state ($n=1$) and of its evolution through a noisy Gaussian channel (as described in the previous section).
Using Williamson's theorem \cite{williamson}, one can show that for the most general single mode Gaussian state the CM can be parameterized as \cite{adam}
\ba
\sigma (\bar n, \theta, r)=(1+2{\bar n})R(\theta)S(2r)R(\theta)^\top.
\label{will}
\ea 
Here the real and positive 
\ba
\bar n :={\mathrm{Tr}}[{\hat a}^\dagger{\hat a}\hat\rho]
\label{nbar}
\ea
is the average occupation number of the Gaussian bosonic mode (e.g., the average photon number),
\ba
R(\theta):=\left (\begin{array}{cc}
\cos\theta ,&\sin\theta \\
-\sin\theta ,& \cos\theta \end{array}\right )
\label{rotmat}
\ea
is an orthogonal  symplectic matrix corresponding to the phase rotation of the mode with angle $\theta\in [0, \pi/4] $, generated by the Hamiltonian $\hat H_\theta= 
\theta {\hat a}^\dagger\hat a$ (giving, in the Heisenberg representation, $\hat a \rightarrow e^{-i\theta}\hat a$), while  
\ba
S(2r):=\left (\begin{array}{cc}
e^{-2r}, & 0\\
0, & e^{2 r} \end{array}\right )
\label{sqmat}
\ea 
is a symplectic transformation corresponding to the squeezing of the mode with parameter $
2r \in {\mathbbm{R}}$, generated by the Hamiltonian $\hat H_r= ir({\hat a}^2-{\hat a}^{\dagger 2})$ (giving, in the Heisenberg representation, $\hat a\rightarrow \hat a\cosh 
 2r ~ - {\hat a}^\dagger\sinh 2r $).

We can also introduce the purity $\mu\in [0, 1]$ of the quantum state in terms of the characteristic function as
\ba
\mu := {\mathrm{Tr}}[{\hat\rho}^2]=\frac{1}{\pi}\int |\chi_\rho(\vec{\xi})|^2d^{2n}\vec{\xi},
\label{purity}
\ea
where in the second step we have used Eq. \rf{scalprod}.
Using Eq. \rf{chig} for a one mode Gaussian quantum state, one easily obtains \cite{marian}
\ba
\mu=\frac{1}{\sqrt{{\mathrm{Det}}\sigma}}=\frac{1}{1+2\bar n},
\label{puritydet}
\ea
where in the second step we have used Eq. \rf{will} and the fact that ${\mathrm{Det}}[R(\theta)]={\mathrm{Det}}[S(2r)]=1$. 
Therefore, one can explicitly rewrite the parametrized CM as
\ba
\sigma =\frac{1}{\mu}
\left (\begin{array}{c}
\cosh 2r -\cos 2\theta\sinh 2r,~\sin 2\theta\sinh 2r\\
 \sin 2\theta\sinh 2r, ~\cosh 2r +\cos 2\theta\sinh 2r  \end{array}\right ).
\label{parcm}
\ea

\section{Free dynamics of one mode Gaussian states in Gaussian channels}

Now let us consider a one mode Gaussian state subject to a noisy Gaussian channel and with Hamiltonian $\hat H=0$.
The dynamics of the CM without the help of external controls is described by Eq. \rf{mastercm}.
After some simple algebra, writing $M:= M_1+iM_2$, it is possible to show that 
\ba
C^\dagger C= \frac{\gamma}{4}\left (\begin{array}{cc}
2N+1 -2M_1,&i  -2M_2\\
-i  -2M_2,&  2N+1 +2M_1
\end{array}\right )
\label{cdagc}
\ea
and therefore, the drift and the diffusion matrices (with $H_0=0$) are respectively given by
\ba
A&=&-\frac{\gamma}{2} I, \label{drift}\\
D&=&\gamma [(2N+1)I +2(M_2\sigma_x +M_1\sigma_z)],
\label{diffusion}
\ea
where $\{\sigma_i ; ~i=x, y, z\}$ are the Pauli matrices. 

The fixed point $\sigma_{\mathrm{fp}}$ of the dissipative dynamics \rf{mastercm} for the CM without the aid of any external control is found by imposing that $\dot\sigma |_{\mathrm{fp}}=0$, and we obtain \cite{paris1}
\ba
\sigma_{\mathrm{fp}}=\frac{D}{\gamma}.
\label{fp}
\ea
Similarly, the fixed point of the dynamics of the displacement vector is given by $\vec{d}_{\mathrm{fp}}=\vec{0}$.
In particular, exploiting Eq. \rf{parcm} and Eqs. (\ref{drift}-\ref{fp}), we obtain
\ba
\mu_{\mathrm{fp}}&=&[(2N+1)^2-4|M|^2]^{-1/2}, \label{muinftycond}\\
\sinh 2r_{\mathrm{fp}}&=&2\mu_{\mathrm{fp}} |M|, \\
\tan 2\theta_{\mathrm{fp}} &=&-\frac{M_2}{M_1},
\label{inftycond}
\ea
where we have $r_{\mathrm{fp}}>0$ and we choose $\theta_{\mathrm{fp}}\in [0, \pi/4]$.
From Eqs. \rf{puritydet} and \rf{muinftycond} one can see that $N$ corresponds to the mean thermal mode number $\bar n$ of the asymptotic Gaussian state only when $M=0$, i.e. in the absence of squeezing. Note that the presence of squeezing $M\not =0$ implies that $\bar n$ is smaller than $N$.

Eq. \rf{fp} also enables to rewrite the Eq. \rf{mastercm}  for the dissipative dynamics of the CM in the more compact form
\ba
\dot\sigma=\gamma(\sigma_{\mathrm{fp}} -\sigma).
\label{mastercm1}
\ea
The latter can be integrated in a straightforward way as \cite{paris1}
\ba
\sigma(t)=e^{-\gamma t}\sigma(0)+(1-e^{-\gamma t})\sigma_{\mathrm{fp}},
\label{cmsol}
\ea
where $\sigma(0)$ represents  the initial correlations of the Gaussian mode.
Clearly, the CM $\sigma(t)$ asymptotically approaches the fixed point of the dynamics without external controls when $t\rightarrow \infty$.
In other words, we have that  $\mu_\infty:=\mu(t\rightarrow\infty)=\mu_{\mathrm{fp}}$, $r_\infty:=r(t\rightarrow\infty)=r_{\mathrm{fp}}$ and
$\theta_\infty:=\theta(t\rightarrow\infty)=\theta_{\mathrm{fp}}$.  

 From Eq.\ \eqref{cmsol} one can see that a convenient parameterization 
of a generic dissipative Gaussian channel of the form \rf{mastercm} can be given in terms of the CM of the corresponding fixed point $\sigma_{\mathrm{fp}}$.
In other words, the triplet 
$(\mu_{\mathrm{fp}}, r_{\mathrm{fp}}, \theta_{\mathrm{fp}})$
completely characterizes the channel. 
After some simple algebra one obtains the following differential equations for the dynamics without control 
of the purity $\mu(t)$, the squeezing $r(t)$ and the phase $\theta(t)$ of the one mode Gaussian state \cite{paris1}
\ba
\dot\mu& =&\gamma\mu \biggl \{1- \frac{\mu}{\mu_{\mathrm{fp}}}[\cosh 2r_{\mathrm{fp}} \cosh 2r \nonumber \\
&-&\cos 2(\theta -\theta_{\mathrm{fp}})\sinh 2r_{\mathrm{fp}}\sinh 2r] \biggr \},\label{dotmu}\\
\dot r& =&-\frac{\gamma\mu}{2\mu_{\mathrm{fp}}} [\cosh 2r_{\mathrm{fp}} \sinh 2r\nonumber \\
&-&\cos 2(\theta -\theta_{\mathrm{fp}})\sinh 2r_{\mathrm{fp}}\cosh 2r], \label{dotr}\\
\dot \theta& =&\frac{\gamma\mu\sinh 2r_{\mathrm{fp}}}{2\mu_{\mathrm{fp}}\cos 2\theta\sinh 2r} [\sin 2 \theta_{\mathrm{fp}} \nonumber \\
&-&\cos 2(\theta -\theta_{\mathrm{fp}})\sin 2 \theta ].
\label{dottheta}
\ea

One can either directly integrate the latter equations or alternatively (and more easily) use the compact solution for $\sigma(t)$, thus obtaining \cite{paris1}
\ba
\mu(t)&=& \mu_0\biggl \{e^{-2\gamma t}  +\frac{2\mu_0}{\mu_{\mathrm{fp}}}[\cosh 2r_0\cosh 2r_{\mathrm{fp}}\nonumber \\
&-&\cos 2(\theta_0-\theta_{\mathrm{fp}})\sinh 2r_0\sinh 2r_{\mathrm{fp}} ]e^{-\gamma t}\nonumber \\
&\times&(1-e^{-\gamma t})+\left(\frac{\mu_0}{\mu_{\mathrm{fp}}}\right )^2 (1-e^{-\gamma t})^2 \biggr \},^{-1/2}\label{musol}\\
\cosh 2r(t) &=& \frac{\mu(t)}{\mu_0} \biggl [ \cosh 2r_0 e^{-\gamma t}\nonumber \\
&+&\frac{\mu_0}{\mu_{\mathrm{fp}}}\cosh 2r_{\mathrm{fp}}(1-e^{-\gamma t})\biggr ],\label{rsol}\\
\tan 2\theta(t)&=& \biggl [ (\sinh 2r_0\sin 2\theta_0e^{-\gamma t} \nonumber \\
&+&\frac{\mu_0}{\mu_{\mathrm{fp}}}\sinh 2r_{\mathrm{fp}}\sin 2\theta_{\mathrm{fp}}(1-e^{-\gamma t})\biggr ]\nonumber \\
&\times&\biggl [ (\sinh 2r_0\cos 2\theta_0e^{-\gamma t} \nonumber \\
&+&\frac{\mu_0}{\mu_{\mathrm{fp}}}\sinh 2r_{\mathrm{fp}}\cos 2\theta_{\mathrm{fp}}(1-e^{-\gamma t})\biggr ]^{-1}
\label{thetasol}
\ea
where $\mu_0:=\mu(0), r_0:=r(0)$ and $\theta_0:=\theta(0)$ are the initial conditions for the purity, squeezing and phase, respectively. 

Furthermore, one can first eliminate the explicit time dependence $\exp[-\gamma t]$ using Eq. \rf{thetasol} and then eliminate the dependence on the phase $\theta$. 
After some elementary but lengthy algebra (see Appendix A), it is then possible to derive an analytical 
formula for the curve $\mu =\mu(r)$,
\ba
\frac{\mu(r)}{\mu_{\mathrm{fp}}}=\frac{a_1\cosh 2r -a_2\sqrt{a_3\sinh^2 2r -a_4}}{a_5},
\label{muofr}
\ea
where the constants $a_i, ~i=1,É5$ are defined in Eq. \rf{a}.

\section{Time optimal control with unconstrained Hamiltonian}

Our main task is to study the time optimal, open-loop, coherent quantum control of the evolution of a one mode Gaussian state under the action of the master equation 
(\ref{masterd}-\ref{mastercm}).
The coherent (unitary) control is now achieved via a Hamiltonian performing phase rotations $\theta(t)$, squeezing $r(t)$ or displacements $\vec{d}(t)$.
We assume that the dissipative part of the quantum evolution~(\ref{dissipator}-\ref{classicalnoise}) is fixed and assigned.  
 We also exclude the possibility of performing measurements on the system to update the quantum control during the evolution, 
 i.e.  no feedback is allowed (notice however that complete information on the initial state of the Gaussian mode
 $(\mu_0, r_0, \theta_0) $ is assumed). 

Within this theoretical framework  we analyze  how to evolve the system  towards a target state ${\hat\rho}_f$, i.e. a state with displacement vector $\vec{d}_f$ 
and CM $\sigma_f$, in the shortest possible time. 
 In more details, we take the target as the fixed point of the dissipative part  of the master equation, i.e. 
a state with $\sigma_f=\sigma_{\mathrm{fp}}$ and $\vec{d}_f=\vec{d}_{\mathrm{fp}}$ fulfilling the condition ${\tilde {\cal{L}}} ({\hat\rho}_{\mathrm{fp}}) =  0$.
This state is a stationary solution (i.e. $\dot{\hat\rho}=0$) of the master equation Eq.~(\ref{master}) when no Hamiltonian is present. 
It represents the attractor points for the dissipative part of evolution, i.e. the states where noise would typically drive the system.
By setting ${\hat\rho}_f= {\hat\rho}_{\mathrm{fp}}$ in our time-optimal analysis we are hence effectively aiming at speeding up relaxation processes that would naturally occur in the system even in the absence of external control. 
In addressing this  issue we do not require perfect unit fidelity, i.e. we tolerate that the quantum state arrives within a  small 
distance from the target, fixed a priori. More precisely, given $\epsilon\in[0,1]$ we look for the minimum value of time $T_{\mathrm{fast}}$ which thanks to a proper choice of $H(t)$   allows us to satisfy
the constraint
\ba
F(\sigma_{\mathrm{opt}}(T_{\mathrm{fast}}), \vec{d}_{\mathrm{opt}}(T_{\mathrm{fast}}); \sigma_{\mathrm{fp}}, \vec{d}_{\mathrm{fp}})=1-\epsilon,
\label{fidcond}
\ea
where the fidelity distance between two one mode Gaussian states with CM $\sigma_1$ and $\sigma_2$ and displacement vectors $\vec{d}_1$ and $\vec{d}_1$ reads \cite{holevo, scutaru}
\ba
F(\sigma_1, \vec{d}_1; \sigma_2, \vec{d}_2):=\frac{2}{\sqrt{\Delta+\delta}-\sqrt{\delta}}e^{-\frac{1}{2}\vec{d}^\top \sigma_+^{-1}\vec{d} },
\label{fidelity}
\ea
with $\Delta :={\mathrm{Det}}(\sigma_1+\sigma_2)$, $\delta :=({\mathrm{Det}}\sigma_1 -1)({\mathrm{Det}}\sigma_2 -1)$, $\vec{d}:=\vec{d}_1-\vec{d}_2$ and $\sigma_+:=\sigma_1+\sigma_2$.

We note here that the dynamics of the displacement vector (Eq. \rf{masterd}) decouples from that of the CM and it can be controlled simply via the use of an external displacement Hamiltonian ${\hat H}_d:=d (e^{i\alpha}\hat a +e^{-i\alpha}{\hat a}^\dagger )$.
In other words, with a proper external control we can instantaneously change the displacement vector from its initial value $\vec{d}_0$ to any desired target.
Effectively, we can thus forget about the dynamics of the displacement vector part of the Gaussian mode and concentrate only on the time optimal control of its CM part.

First of all we compute the minimal time $T_{\mathrm{free}}(\sigma_0,\epsilon)$ required for an initial state $\sigma_0$  
 to freely reach the target $\sigma_{\mathrm{fp}}$ within a fixed fidelity $1-\epsilon$  under the sole action of decoherence and without the aid of any external control.
We derive $T_{\mathrm{free}}$ from Eq. \rf{musol} evaluated at $t=T_{\mathrm{free}}$ and where 
$\mu(T_{\mathrm{free}}):=\mu_{\mathrm{Tf}}$ is found by imposing the fidelity condition (\ref{fidcond}-\ref{fidelity}) with $\vec{d}=0$. 
From Eq. \rf{parcm}, we compute
\ba
\Delta_{\mathrm{Tf}}&=&\frac{1}{\mu^2_{\mathrm{Tf}}}\biggl [1+2\frac{\mu_{\mathrm{Tf}}}{\mu_{\mathrm{fp}}}(\cosh 2r_{\mathrm{Tf}}\cosh 2r_{\mathrm{fp}}
\nonumber\\
&-& 
\cos 2(\theta_{\mathrm{Tf}}-\theta_{\mathrm{fp}})\sinh 2r_{\mathrm{Tf}}\sinh 2r_{\mathrm{fp}})+\frac{\mu^2_{\mathrm{Tf}}}
{\mu^2_{\mathrm{fp}}}\biggr ]
\label{Delta}\\
\delta_{\mathrm{Tf}}&=&\frac{(1-\mu^2_{\mathrm{fp}})}{\mu^2_{\mathrm{fp}}\mu^2_{\mathrm{Tf}}}(1-\mu^2_{\mathrm{Tf}}),
\label{delta}
\ea 
where we have defined $\Delta_{\mathrm{Tf}}:=\Delta(T_{\mathrm{free}})$,
$\delta_{\mathrm{Tf}}:= \delta(T_{\mathrm{free}})$, $r_{\mathrm{Tf}}:=r(T_{\mathrm{free}})$ and $\theta_{\mathrm{Tf}}:=\theta(T_{\mathrm{free}})$.
Now, for a generic $\sigma_0$ not close to the target, we will have $\gamma T_{\mathrm{free}}>1$ and therefore we can 
obtain $\mu_{\mathrm{Tf}}, r_{\mathrm{Tf}}$ and $\theta_{\mathrm{Tf}}$ via an expansion in 
$\exp[-\gamma T_{\mathrm{free}}]$ of Eqs. \rf{musol}, \rf{rsol} and \rf{thetasol}, respectively.
After a lengthy but elementary algebraic manipulation, upon imposing conditions (\ref{fidcond}-\ref{fidelity}) one finally finds
\ba
T_{\mathrm{free}}(\sigma_0,\epsilon)
&=&\frac{1}{\gamma}\ln \{
[(1+\mu^2_{\mathrm{fp}})(\mu_0-\mu_{\mathrm{fp}}\beta_0)^2\nonumber\\
&+&\mu^2_{\mathrm{fp}}(1-\mu^2_{\mathrm{fp}})
(\beta_0^2-1)]^{1/2}\nonumber\\
&\times &[2\mu_0\sqrt{\epsilon}\sqrt{1-\mu^4_{\mathrm{fp}}}]^{-1}\}\simeq \frac{|\ln\epsilon|}{2\gamma},
\label{Tfree}
\ea
where  
we have introduced
$\beta_0:=\cosh 2r_0\cosh 2r_{\mathrm{fp}} - \cos 2(\theta_0-\theta_{\mathrm{fp}}) \sinh 2r_0\sinh 2r_{\mathrm{fp}}$.
As expected, for a generic initial state $T_{\mathrm{free}}$ diverges as $\epsilon \rightarrow 0$.
This function sets the benchmark that we will use to compare the performance of our time-optimal control problem.

Next, we address  the problem of speeding up the transition of the system from $\sigma_0$ towards
the fixed point  state $\sigma_{\mathrm{fp}}$ with a proper engineering of the quantum control Hamiltonian $H(t)$ to
see how much one can gain with respect to the  ``natural" time  $T_{\mathrm{free}}$
of Eq.~(\ref{Tfree}). Clearly the result will depend strongly on the freedom we have in choosing the functions $r(t), \theta(t)$ and $d(t)$.

For a coherent control where the choice  of the possible functions $r(t), \theta(t)$ and $d(t)$ is unconstrained 
the problem  essentially reduces to finding the maximum of the modulus of the speed of purity change, at any given purity, for the Gaussian channel.
As we have already said when computing $T_{\mathrm{free}}$, with a proper external control we can instantaneously change the displacement vector from its initial value $\vec{d}_0$ to any desired target and thus, 
effectively, we can forget about the dynamics of the displacement vector.
Then, given any arbitrary initial CM  $\sigma_0$ of the one mode Gaussian state,
 one can always unitarily and instantaneously (since we may take a control with infinite strength) move from the initial point along the surface of constant purity  $\mu_0$ until one reaches the new position of coordinates $(\mu_0, r_{\mathrm{ext}}, \theta_{\mathrm{ext}})$ where the speed of purity change induced by the dissipator, i.e. 
the quantity
\ba
v(r, \theta ):= \dot\mu,
\label{purity}
\ea
is extremal for fixed purity $\mu_0$, where $\dot\mu$ is given by Eq. \rf{dotmu}. 
Then, one can switch off the control and let the system decohere for a time $T_{\mathrm{fast}}$ until the purity  $\mu(T_{\mathrm{fast}})$ 
which satisfies the fidelity condition (\ref{fidcond}) is reached.
Finally, one can switch the quantum control on again and unitarily rotate the CM from
the position $(\mu(T_{\mathrm{fast}}), r_{\mathrm{ext}}, \theta_{\mathrm{ext}})$ to a point within tolerable distance from the target at 
$(\mu(T_{\mathrm{fast}}), r_{\mathrm{fp}}, \theta_{\mathrm{fp}})$.

In the following, we assume that we are able to perform instantaneous squeezing of the Gaussian mode up to a maximum strength, i.e.  we take $|r|< r_M$, with $r_M \gg 1$).

\subsection{Speeding up the relaxation}

There are two possible scenarios to consider: a) the cooling case, i.e. when $\mu_0<\mu_{\mathrm{fp}}$, and b) the heating case, i.e. when $\mu_0>\mu_{\mathrm{fp}}$.
Since we assume that unitary operations can be done arbitrarily fast, i.e. that $r$ and $\theta$ can be changed instantaneously, the dynamics is effectively captured 
by Eqs. \rf{dotmu} and \rf{purity} for the speed of change of the purity at a given purity.

The optimal values of the speed $v(r, \theta)$ at given $\mu$ are found by imposing that $\nabla v(r, \theta)=0$.
In particular, a set of locally positive maxima of the speed $v$ is obtained for $\mu \not =0$ and 
\ba
r_{\mathrm{ext, M}}&=&r_{\mathrm{fp}},\label{crit1r}\\
\theta_{\mathrm{ext, M}} &=& \theta_{\mathrm{fp}}.
\label{crit1t}
\ea
In this case the speed of purity change \rf{dotmu} reads $v=\gamma\mu(1-\mu/\mu_{\mathrm{fp}})$.
Moreover, we have that $v$ has other local stationary points on the boundary of the domain of the allowed parameters, i.e. along the curves
of maximum squeezing $r=\pm r_M$ or with  $\theta =0$.
In particular, local maxima of $v$ are attained at $(r_{\mathrm{ext}} \theta_{\mathrm{ext}})=(r_M, \theta_{\mathrm{fp}})$, where we have 
$v=\gamma\mu [1-(\mu/\mu_{\mathrm{fp}})\cosh 2(r_M-r_{\mathrm{fp}})]$, while global minima of $v$ are attained at 
\ba
r_{\mathrm{ext, m}}&=&-r_M, \label{crit2r}\\
\theta_{\mathrm{ext, m}}&=&\theta_{\mathrm{fp}},
\label{crit2t}
\ea
where we have $v=\gamma\mu [1-(\mu/\mu_{\mathrm{fp}})\cosh 2(r_M+r_{\mathrm{fp}})]$.
Another local maximum if found at $\theta=0$ and $\tanh 2r =\cos 2\theta_{\mathrm{fp}}\tanh 2 r_{\mathrm{fp}}$, for which we have
$v=\gamma\mu [1-\mu/\mu_{\mathrm{fp}}(1+\sin^22\theta_{\mathrm{fp}}\sinh^2 2r_{\mathrm{fp}})^{1/2}]$.

Therefore, in the case a) of cooling, i.e. when we want to reach $\mu_{\mathrm{fp}}$ starting from $\mu_0 <\mu_{\mathrm{fp}}$ and we look for a maximum of $v$, the optimal solution $v^{\mathrm{cool}}_{\mathrm{fast}}$ is obtained at the critical point \rf{crit1r}-\rf{crit1t} and is given by
\ba
v^{\mathrm{cool}}_{\mathrm{fast}}=\gamma\mu\biggl (1-\frac{\mu}{\mu_{\mathrm{fp}}}\biggr )>0.
\label{vcool}
\ea
Furthermore, from Eqs. (\ref{dotr}-\ref{dottheta}) we note that at the global 
maximum
of the  speed $v$ the values of $r$ and 
$\theta$ are stationary, i.e. $\dot r=\dot\theta=0$.
Then, the optimal control strategy is the following (Fig. 1a): \\
i) rotate from $\theta_0$ to $\theta_{\mathrm{fp}}$ and squeeze from $r_0$ to $r_{\mathrm{fp}}$; \\
ii) let the Gaussian mode freely decohere from $\mu_0$ to $\mu_{\mathrm{T c}}$ (with the external controls off and
with $r$ and $\theta$ stationary at $r_{\mathrm{fp}}$ and $\theta_{\mathrm{fp}}$, respectively).\\

The optimal time to let the system cool from the initial state $(\mu_0, r_0, \theta_0)$ to the fixed point $(\mu_{\mathrm{fp}}, r_{\mathrm{fp}}, \theta_{\mathrm{fp}})$ is thus given by the formula
\ba
T^{\mathrm{cool}}_{\mathrm{fast}}(\sigma_0,\epsilon)
=\int_{\mu_0}^{\mu_{\mathrm{T c}}} \frac{d\mu}{v^{\mathrm{cool}}_{\mathrm{fast}}}.
\label{tcooldef}
\ea
The purity $\mu_{\mathrm{T c}}:=\mu(T^{\mathrm{cool}}_{\mathrm{fast}})$ is computed along the lines described in the previous Section for the case of the free decoherence dynamics.
One evaluates the quantities $\Delta$ and $\delta$ and then the fidelity $F$ in Eqs. (\ref{fidcond}-\ref{fidelity}) for $\vec{d}=0$, with 
$\sigma_1=\sigma(\mu_{\mathrm{T c}}, r_{\mathrm{fp}}, \theta_{\mathrm{fp}})$ and 
$\sigma_2=\sigma_{\mathrm{fp}}$, obtaining
\ba
\mu_{\mathrm{T c}}\simeq \mu_{\mathrm{fp}}(1-2\sqrt{\epsilon}\sqrt{1-\mu^2_{\mathrm{fp}}}).
\label{mucool}
\ea 
for all mixed fixed points $\mu_{\mathrm fp}\neq 1$ (the special case of $\mu_{\mathrm{fp}}=1$ is considered in Appendix B).

Then the integral in \rf{tcooldef} can be computed explicitly upon using \rf{vcool} and \rf{mucool}, and we get
\ba
T^{\mathrm{cool}}_{\mathrm{fast}}(\sigma_0,\epsilon)
=\frac{1}{\gamma}\ln \biggl [\frac{(\mu_{\mathrm{fp}} -\mu_0)\mu_{\mathrm{T c}}}{(\mu_{\mathrm{fp}} -\mu_{\mathrm{T c}})\mu_0}\biggr ]\simeq \frac{|\ln \epsilon |}{2\gamma},
\label{tcoolexp}
\ea
which diverges as $\epsilon \rightarrow 0$.

%%%%%%%%%%%%%%%%%%%%%%%%%%%%%%%%%%%%%%%%%%%%%
\begin{figure}[ht]
\begin{center}
\hspace*{-1cm}
\includegraphics[scale=0.4,angle=0]{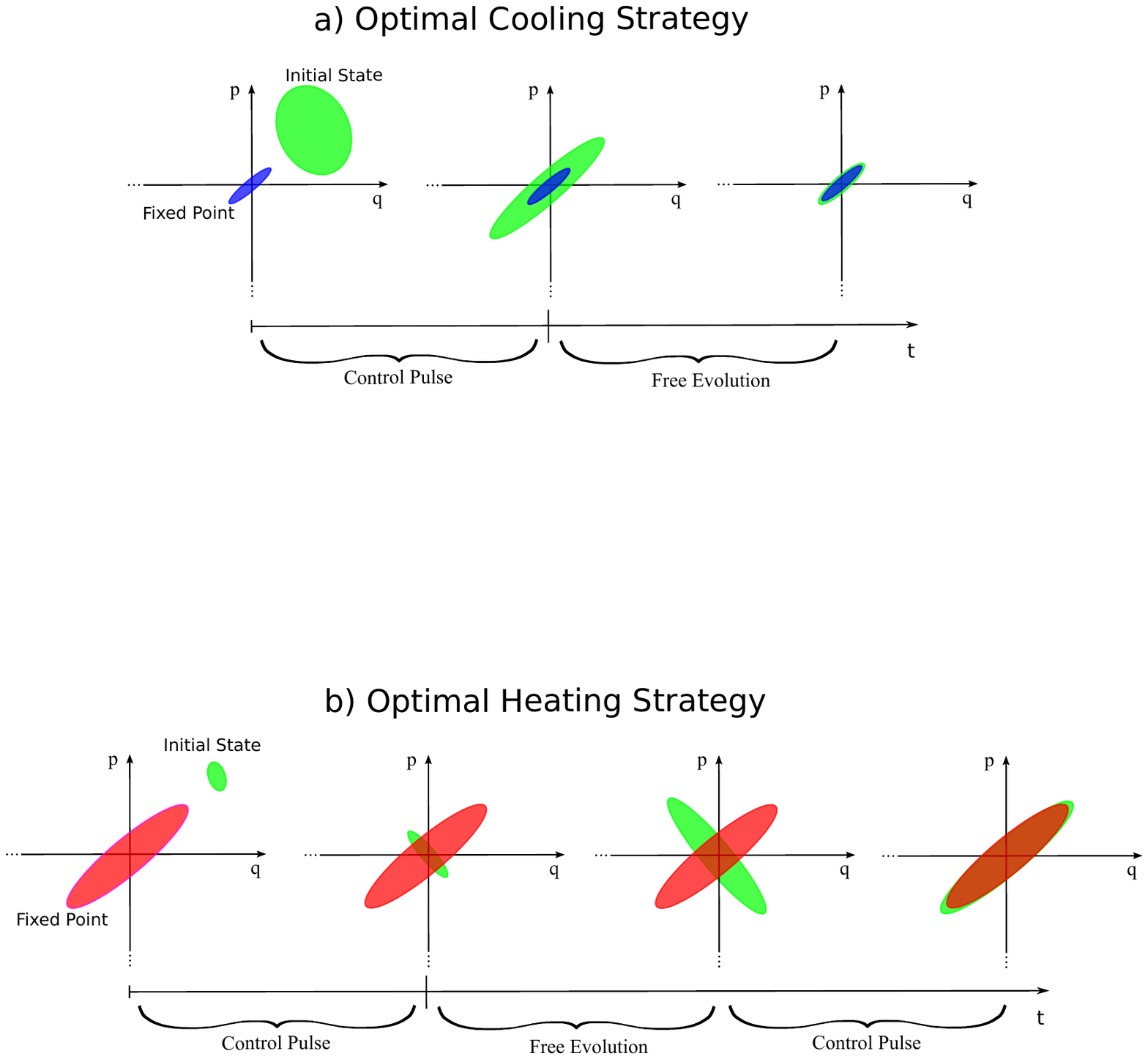}
\end{center}
\caption {Time optimal cooling (a) and heating (b) strategies for a single mode Gaussian state in a Markovian dissipative channel when instantaneous unitary control is available.}
\label{newfig}
\end{figure}
%%%%%%%%%%%%%%%%%%%%%%%%%%%%%%%%%%%%%%%%%%%%%%

In the heating case b), i.e. when we want to reach $\mu_{\mathrm{fp}}$ starting from  $\mu_0>\mu_{\mathrm{fp}}$ and we look for a minimum of $v$, the optimal solution is  $v^{\mathrm{heat}}_{\mathrm{fast}}$ is obtained along the boundary at the point specified by \rf{crit2r}-\rf{crit2t}
and it reads
\ba
v^{\mathrm{heat}}_{\mathrm{fast}}=\gamma\mu\biggl [1-\frac{\mu}{\mu_{\mathrm{fp}}}\cosh 2(r_M+r_{\mathrm{fp}})\biggr ]<0.
\label{vheat}
\ea
In this case, the optimal control strategy consists in (Fig. 1b): \\
i) rotating from $\theta_0$ to $\theta_{\mathrm{fp}}$ and squeeze from $r_0$ to $-r_M$;\\
ii) letting the mode 
freely
decohere from $\mu_0$ to $\mu_{\mathrm{T h}}$ while keeping $r=-r_M$ fixed ($\dot r=0$) via an appropriate squeezing Hamiltonian \footnote{Note that also in this case, as for cooling, $\theta$ is stationary at the critical point without the need of any Hamiltonian control.}; \\
iii) squeezing from $-r_M$ to $r_{\mathrm{fp}}$. 
\\

The optimal time to 
heat up
 the system from the initial state $(\mu_0, r_0, \theta_0)$ to the fixed point $(\mu_{\mathrm{fp}}, r_{\mathrm{fp}}, \theta_{\mathrm{fp}})$
is thus given by
\ba
T^{\mathrm{heat}}_{\mathrm{fast}}(\sigma_0,\epsilon)
=\int_{\mu_0}^{\mu_{\mathrm{T h}}} \frac{d\mu}{v^{\mathrm{heat}}_{\mathrm{fast}}}.
\label{theatdefa}
\ea
The computation of the purity $\mu_{\mathrm{T h}}:=\mu(T^{\mathrm{heat}}_{\mathrm{fast}})$ is similar to that done for the
cooling case, where now in Eqs. (\ref{fidcond}-\ref{fidelity}) we use $\sigma_1=\sigma(\mu_{\mathrm{T h}}, -r_{\mathrm{M}}, \theta_{\mathrm{fp}})$ and $\sigma_2=\sigma_{\mathrm{fp}}$, and we obtain
\ba
\mu_{\mathrm{T h}}\simeq \mu_{\mathrm{fp}}(1+2\sqrt{\epsilon}\sqrt{1-\mu^2_{\mathrm{fp}}}).
\label{muheat}
\ea 
Then the integral in Eq. \rf{theatdefa} can be evaluated explicitly upon using \rf{vheat} and \rf{muheat}, and we get
\ba
T^{\mathrm{heat}}_{\mathrm{fast}}(\sigma_0,\epsilon)
&=&\frac{1}{\gamma}\ln \biggl \{\frac{[\mu_{\mathrm{fp}} -\mu_0\cosh 2(r_M+r_{\mathrm{fp}})]\mu_{\mathrm{T h}}}{[\mu_{\mathrm{fp}} -\mu_{\mathrm{T h}}\cosh 2(r_M+r_{\mathrm{fp}})]\mu_0}\biggr \}\nonumber \\
&\simeq&\frac{1}{\gamma}\log \biggl \{ \frac{[\mu_0\cosh 2(r_M+r_{\mathrm{fp}})-\mu_{\mathrm{fp}}]}{\mu_0[\cosh 2(r_M+r_{\mathrm{fp}})-1)}\biggr \}
\label{theatsola}
\ea
which is finite.

%{\color{blue}
Finally, we can compare the relaxation time of the quantum controlled dynamics with the 
relaxation time of the free evolution for a single bosonic mode in contact with a generic environment.
We start with an important case which is worth to be considered separately because of its potential implications in quantum thermodynamics, namely that of an initial Gibbs state ($r_0=0$) placed in contact with a non-squeezed thermal bath ($r_{\mathrm{fp}}=0$) characterized by a different temperature ($\mu_0\neq\mu_{\mathrm{fp}}$). In this case our results suggest that it is impossible to increase the cooling rate of a single Gaussian mode by quantum control, while it is possible to increase the heating rate. This fact is analogue to the time optimal control of a single qubit  \cite{mukherjee}. 

For a generic initial state and a generic dissipative channel, unless the initial state is already along the extremal trajectory with $r_0=r_{\mathrm{fp}}$ and $\theta_0=\theta_{\mathrm{fp}}$,
quantum control is always advantageous for speeding up the relaxation process. However, in the limit of a small error parameter $\epsilon \rightarrow 0$, the situation is very
different depending on the 
%{\color{red} 
relative values of the purity of the initial state and of the fixed point.
%}  
In the cooling case $\mu_0 < \mu_{\mathrm{fp}}$, comparing Eqs. \eqref{Tfree} and \eqref{tcoolexp}, we see that in the limit $\epsilon \rightarrow 0$  the optimal time is asymptotically
equivalent  to the free evolution time, since both quantities diverge as $|\ln(\epsilon)|/(2\gamma)$ (or as $|\ln(\epsilon)|/\gamma$ in the special case $\mu_{\mathrm{fp}}=1$ as explained in Appendix B). In the heating case $\mu_0 > \mu_{\mathrm{fp}}$ instead, the optimized relaxation time is dramatically different from the free evolution time.  Indeed, from Eq.\ \eqref{theatsola} we observe that the
fixed point can be reached exactly ($\epsilon=0$) in a finite time, while the free evolution time diverges to infinity.
%}{\color{red} 
We can also introduce a measure of the performance of the quantum control in the worst case scenario by maximizing the time durations of the evolutions with respect to the possible initial states of the Gaussian mode, in a way similar to what done in our previous work on the discrete model case \cite{mukherjee}.
In fact, one can show that $T_{\mathrm{free}}(\sigma_0,\epsilon)$ in Eq. \rf{Tfree} is maximum for 
$\mu_0/\mu_{\mathrm{fp}} \rightarrow 0$ and $|r_0|\rightarrow \infty$, and for $\epsilon \rightarrow 0$ we obtain that the longest time one would have to wait to bring the Gaussian mode close to the fixed point in the absence of external controls goes like
\ba
T^{\mathrm{max}}_{\mathrm{free}}&:=&\max_{r_0, \mu_0, \theta_0}T_{\mathrm{free}}(\sigma_0,\epsilon)\nonumber \\
&\simeq& \frac{1}{\gamma}
\lim_{\stackrel{\epsilon, \frac{\mu_0}{\mu_{\mathrm{fp}}} \to 0}{|r_0|\to \infty}} 
\left [\biggl |\ln \left (\frac{\mu_0\sqrt{\epsilon}}{\mu_{\mathrm{fp}}}\right )\biggr |+2|r_0|\right ],
\label{Tfreemax}
\ea
where we have used the fact that $2\beta_0\rightarrow \exp[2|r_0|]$ for $|r_0|\rightarrow \infty$.
On the other hand, by studying the optimal times to cool and heat the system, Eqs. \rf{tcoolexp} and \rf{theatsola} one can easily check that the optimal time one would have to wait to reach the target with the help of unconstrained control is also achieved for  $\mu_0/\mu_{\mathrm{fp}} \rightarrow 0$ and, for $\epsilon \rightarrow 0$, this goes like
\ba
T^{\mathrm{max}}_{\mathrm{fast}}&:=&\max_{r_0, \mu_0, \theta_0}T_{\mathrm{fast}}(\sigma_0,\epsilon)
\nonumber \\
&\simeq& \frac{1}{\gamma}
\lim_{\stackrel{\epsilon, \frac{\mu_0}{\mu_{\mathrm{fp}}} \to 0}{|r_0|\to \infty}} 
\biggl |\ln \left (\frac{\mu_0\sqrt{\epsilon}}{\mu_{\mathrm{fp}}}\right )\biggr |.
\label{tcoolexpmax}
\ea
By comparing the maxima \rf{Tfreemax} and \rf{tcoolexpmax} we conclude that quantum control enhances the performance by a factor
\ba
\frac{T^{\mathrm{max}}_{\mathrm{free}}}{T^{\mathrm{max}}_{\mathrm{fast}}}
\simeq 1+\lim_{\stackrel{\epsilon, \frac{\mu_0}{\mu_{\mathrm{fp}}} \to 0}{|r_0|\to \infty}} 
 \frac{2|r_0|}{\biggl |\ln\left(\frac{\mu_0\sqrt{\epsilon}}{\mu_{\mathrm{fp}}}\right )\biggr |},
\label{performancefactor}
\ea
which is much larger than one if $2|r_0| \gg | \ln (\mu_0\sqrt{\epsilon} / \mu_{\mathrm{fp}} )|$.
%}

\subsection{
%{\color{blue} 
Stopping
%}
 the relaxation }

%{\color{blue}
In the previous analysis we focused on the task of speeding up the relaxation of an open system via quantum control. Sometimes, however, one may be interested in the opposite task of stopping the dissipative dynamics and avoiding the natural evolution of the state towards the fixed point in equilibrium with the environment.

The dynamics induced by the dissipative channel is described by the differential equations (\ref{dotmu}), (\ref{dotr}) and (\ref{dottheta}). In the 
%}{\color{red} 
heating
%} {\color{blue} 
case $\mu_0>\mu_{\mathrm{fp}}$, one can easily check that $\dot \mu$ is always strictly negative for every $r$ and $\theta$ meaning that decoherence cannot be stopped by quantum control.
Instead, in the cooling case $\mu_0<\mu_{\mathrm{fp}}$, one may have $\dot \mu=0$ for some particular values of $r$ and $\theta$. More precisely, this is achieved for all the states whose parameters $\mu$, $r$ and $\theta$ satisfy the condition  
\ba
\frac{\mu}{\mu_{\mathrm{fp}}} &=& [\cosh 2r_{\mathrm{fp}}\cosh 2r\nonumber \\
& &-\cos 2(\theta -\theta_{\mathrm{fp}})\sinh 2r_{\mathrm{fp}}\sinh 2r)]^{-1}.
\label{vpos}
\ea
In the assumption of unconstrained control, one can always keep $\dot r=0$ and $\dot \theta=0$ by appropriate squeezing and phase shift Hamiltonians. Therefore Eq. \rf{vpos} essentially defines an extended set of `artificial' fixed points, {\it i.e.}\ states which can be forced to remain stationary with the aid of quantum control despite they are not in equilibrium with the environment.  
%}

\section{Discussion}
%{\color{blue}
In this work we studied the time optimal control of a single-mode Gaussian state evolving according to a generic Markovian dissipative master equation.
We focused on the specific task of minimizing the total time necessary for a given initial state to converge close to the fixed point of the dynamics within a given 
error parameter $\epsilon$.

We first computed such relaxation time in the absence of any control field. In this case the system is only driven by decoherence and dissipation until 
the associated quantum state converges to a unique Gaussian state in equilibrium with the environment. Then we optimized such relaxation time assuming that
one can apply arbitrary Gaussian unitary operations to the state during its time evolution. We obtained several analytical results which strongly depend on the initial state of the system and on the
%} {\color{red}
relative values of the purities of the initial state and of the fixed point.
%}{\color{blue}  
In particular, what we found is that in the limit of vanishing error $\epsilon \rightarrow \infty$ the advantage of quantum control is negligible if the purity of the initial state is less than the purity of the fixed point while, in the opposite case, quantum control allows an exponential advantage with respect to the uncontrolled dynamics.
%}{\color{red} 
We further introduced a measure of the performance of the quantum control in the worst case scenario by maximizing the time durations of the evolutions with respect to all possible initial states of the Gaussian mode, and
we found that quantum control greatly enhances the performance whenever 
$2|r_0| \gg | \ln (\mu_0\sqrt{\epsilon} / \mu_{\mathrm{fp}} )|$ for large $r_0$, $\mu_0/\mu_{\mathrm{fp}} \rightarrow 0$ and 
$\epsilon \rightarrow 0$.

%}{\color{blue} 
Our results could find applications in any physical system in which Gaussian unitary operations can be applied sufficiently quickly with respect to the natural decoherence time.
For example in optical and microwave systems with sufficiently large and switchable non-linearities, as typical in experiments of electromagnetically induced transparency \cite{EIT1,EIT2}.
Another potential application could arise in the field of optically levitated nano-spheres \cite{sphere1,sphere2,sphere3}. The motion of such particles is essentially harmonic but one can easily change the trapping potential realizing effective squeezing operations and phase-space rotations. The manipulation of the optical potential would generate a fast and precise control of the state of the trapped particle. 

Finally we would like to stress that our results present also some fundamental aspects which may be interesting from the point of view of quantum thermodynamics \cite{QThermo}. For example, according to our analysis it is impossible to speed up the cooling process of a thermal state in a cold bath, in the standard scenario in which the state and the bath are not squeezed. It would be interesting to investigate how this fact depends on the specific model or if it is a more general property of thermalization processes. 

Other research directions could be: generalizing the target state of time optimal control to arbitrary Gaussian states (not necessarily the fixed point),  extending this approach to multimode bosonic systems, or including the possibility of non-Gaussian operations in the control strategy.

\section{Acknowledgements}
This work was supported by MIUR-FIRB-IDEAS project RBID08B3FM and by GR13(Mari) SNS.

\appendix

\section{Explicit formula for the trajectories}

We eliminate the explicit time dependence $\exp[-\gamma t]$ using Eq. \rf{thetasol} and we get
\ba
e^{-\gamma t}&=& \biggl [ 1 - \frac{\mu_{\mathrm{fp}}\sin 2(\theta -\theta_0)\sinh 2r_0}{\mu_0\sin 2(\theta -\theta_{\mathrm{fp}} )\sinh 2r_{\mathrm{fp}}} \biggr ]^{-1}.
\label{toftheta}
\ea
Then, substituting the latter into Eqs. \rf{musol} and \rf{rsol}, we parameterize the quantum trajectories in terms of the phase $\theta$ as
\ba
\sqrt{c}~\mu(\theta)&=&|\mu_{\mathrm{fp}}\sin 2(\theta -\theta_0)\sinh 2r_0\nonumber \\
&-&\mu_0\sin 2(\theta -\theta_{\mathrm{fp}} )\sinh 2r_{\mathrm{fp}} |,\label{muoftheta}\\
s\sqrt{c}~\cosh 2r(\theta) &=&[\sin 2(\theta -\theta_0)\sinh 2r_0\cosh 2r_{\mathrm{fp}} \nonumber \\
&-&\sin 2(\theta -\theta_{\mathrm{fp}})\cosh 2r_0\sinh 2r_{\mathrm{fp}}],
\label{roftheta}
\ea
where we have defined the constants
\ba
b&:=& \cosh 2r_0\cosh 2r_{\mathrm{fp}} -\cos 2(\theta_0 -\theta_{\mathrm{fp}})\sinh 2r_0\sinh 2r_{\mathrm{fp}},\nonumber \\
c&:=& \sin^2 2(\theta -\theta_0)\sinh^22r_0 + \sin^2 2(\theta -\theta_{\mathrm{fp}})\sinh^22r_{\mathrm{fp}},\nonumber \\
&-& 2b\sin 2(\theta -\theta_0)\sin 2 (\theta -\theta_{\mathrm{fp}})\sinh 2r_0\sinh 2r_{\mathrm{fp}},\nonumber \\
s&:&={\mathrm{sign}}[\mu_{\mathrm{fp}}\sin 2(\theta -\theta_0 )\sinh 2r_0 \nonumber \\
&-&\mu_0\sin 2(\theta -\theta_{\mathrm{fp}})\sinh 2r_{\mathrm{fp}}].
\label{bcs}
\ea
After a lengthy but elementary algebra, it is also possible to eliminate the dependence on the phase $\theta$ and to derive the analytical 
formula for the curve $\mu =\mu(r)$ given by \rf{muofr}, where we have introduced the constants
\ba
a_1&=&\sinh 2r_0[\sinh 2r_0\cosh 2r_{\mathrm{fp}} \nonumber \\
&-&\cos 2(\theta_0-\theta_{\mathrm{fp}})\cosh 2r_0\sinh 2r_{\mathrm{fp}}] \nonumber \\
&+&\frac{\mu_0}{\mu_{\mathrm{fp}}}\sinh 2r_{\mathrm{fp}}[\cosh 2r_0\sinh 2r_{\mathrm{fp}} \nonumber \\
&-&\cos 2(\theta_0-\theta_{\mathrm{fp}})\sinh 2r_0\cosh 2r_{\mathrm{fp}}] ,\nonumber \\
a_2&=&\cosh 2r_0-\frac{\mu_0}{\mu_{\mathrm{fp}}}\cosh 2r_{\mathrm{fp}}, \nonumber \\
a_3&=&-1+[\cosh 2r_0\cosh 2r_{\mathrm{fp}} \nonumber \\
&-&\cos 2(\theta_0-\theta_{\mathrm{fp}})\sinh 2r_0\sinh 2r_{\mathrm{fp}}]^2,\nonumber \\
a_4&=&[\sin 2(\theta_0-\theta_{\mathrm{fp}})\sinh 2r_0\sinh 2r_{\mathrm{fp}}]^2,\nonumber \\  
a_5&=&\sinh^2 2(r_0-r_{\mathrm{fp}}) \nonumber \\
&+&\sin^2 2(\theta_0-\theta_{\mathrm{fp}})\sinh 4r_0\sinh 4r_{\mathrm{fp}}.
\label{a} 
\ea

In particular, the analytical expression $\mu=\mu(r)$ for the curve of the quantum evolution under free decoherence at the optimal point $\theta =\theta_0=\theta_{\mathrm{fp}}$ is given by 
\ba
\mu(r)=\mu_{\mathrm{fp}}\frac{\left [\sinh 2(r+r_0)-\frac{\mu_0}{\mu_{\mathrm{fp}}}\sinh 2(r+r_{\mathrm{fp}})\right ]}
{\sinh2(r_0-r_{\mathrm{fp}})}.
\label{muofropt}
\ea

\section{Pure fixed point}
When the target state, i.e. the fixed point of the free evolution, is given by a pure state, we have $\mu_{\mathrm{fp}}=1$.
In this case we have that $\delta =0$ and therefore $F=2/\sqrt{\Delta}$ (at $\vec{d}=\vec{0}$) in Eqs. (\ref{fidcond}-\ref{fidelity}).
Therefore, for the dynamics of free decoherence from an arbitrary initial point $\sigma_0$ to the target 
$\sigma_{\mathrm{fp}}$ we get, using methods similar to those explained in the main text,
\ba
T_{\mathrm{free, pure}}=\frac{1}{\gamma}\ln\left [\frac{2\mu_0}{(\beta_0-\mu_0)\epsilon}\right ]
\simeq \frac{|\ln \epsilon |}{\gamma},
\label{tfreepure}
\ea
which diverges as $\epsilon \rightarrow 0$.
Since the initial state must have $\mu_0<\mu_{\mathrm{fp}}$, there is just the possibility of optimal cooling now.
The strategy is the same as that of the main test for cooling, where now, however, from the fidelity conditions 
(\ref{fidcond}-\ref{fidelity}) we find
\ba
\mu_{\mathrm{T c, pure}}\simeq 1-2\epsilon.
\label{mucoolpure}
\ea
Therefore, inserting \rf{mucoolpure} into Eq. \rf{tcooldef} we obtain the time optimal cooling time
\ba
T_{\mathrm{fast, pure}}^{\mathrm{cool}}=\frac{1}{\gamma}\ln\left [\frac{(1-\mu_0)}{2\mu_0\epsilon}\right ]
\simeq \frac{|\ln \epsilon |}{\gamma},
\label{tcoolpure}
\ea
which diverges as $\epsilon \rightarrow 0$ 
%{\color{red}  
in the same way as $T_{\mathrm{free, pure}}$.
Therefore, there is no advantage in using time-optimal quantum control for the relaxation towards a pure fixed point. 
%}

\end{document}